\documentclass[twocolumn,showpacs,preprintnumbers,amsmath,amssymb,superscriptaddress]{revtex4}

\usepackage{graphicx}

\usepackage{dcolumn}
\usepackage{bm}
\usepackage{amsmath}
\usepackage{color}

\begin{document}

\author{William N. Plick}\email{billplick@gmail.com}\altaffiliation{Current Address: LTCI CNRS - T\'{e}l\'{e}com ParisTech, Paris, France}

\affiliation{Quantum Optics, Quantum Nanophysics, Quantum Information, University of Vienna, Boltzmanngasse 5, Vienna A-1090, Austria}
\affiliation{Institute for Quantum Optics and Quantum Information, Boltzmanngasse 3, Vienna A-1090, Austria}

\author{Robert Fickler}
\affiliation{Quantum Optics, Quantum Nanophysics, Quantum Information, University of Vienna, Boltzmanngasse 5, Vienna A-1090, Austria}
\affiliation{Institute for Quantum Optics and Quantum Information, Boltzmanngasse 3, Vienna A-1090, Austria}

\author{Radek Lapkiewicz}

\affiliation{Quantum Optics, Quantum Nanophysics, Quantum Information, University of Vienna, Boltzmanngasse 5, Vienna A-1090, Austria}
\affiliation{Institute for Quantum Optics and Quantum Information, Boltzmanngasse 3, Vienna A-1090, Austria}

\author{Sven Ramelow}\altaffiliation{Current Address: Cornell University, 271 Clark Hall, 142 Science Dr., Ithaca, 14853 NY, USA}

\affiliation{Quantum Optics, Quantum Nanophysics, Quantum Information, University of Vienna, Boltzmanngasse 5, Vienna A-1090, Austria}
\affiliation{Institute for Quantum Optics and Quantum Information, Boltzmanngasse 3, Vienna A-1090, Austria}

\title{Violation of a new Wigner inequality with high angular momenta}

\date{\today}


\begin{abstract}
\noindent  In the study of systems that cannot be described classically, the Wigner inequality, has received only a small amount of attention. In this paper we extend the Wigner inequality \--- originally derived in 1969 \--- and show how it may be used to contradict local realism with only coincidence detections in the absence of two-outcome measurements \--- that is, for any system where only one possible result of a pair of potential outcomes can be registered. It thus encapsulates a much broader class of measurement schemes than could previously violate a local-realistic inequality. This is possible due to an assumption of ``extended fairness'' on the measurement outcomes, which we posit is highly plausible. We then apply this inequality to a recently constructed setup with access to entangled pairs of photons with very high angular momenta, in which no previously derived local-realistic inequality could successfully be used without making very broad assumptions. To our knowledge this experiment constitutes a violation of local realism with the largest quanta to date. We thus demonstrate the versatility of this inequality under very lossy conditions.
\end{abstract}

\maketitle

\section{Introduction}

\noindent The ability to banish the local-realistic world-view can be an important benchmark for experiments in quantum physics. Much effort has been dedicated in recent decades to highlight the contradictory predictions of classical and quantum mechanics with ever larger and less conventional systems. As a result, in order to explain recent results realistically, one must believe an increasingly unlikely set of assumptions; as the number and complexity of setups has increased dramatically. At the forefront of those efforts are the experiments which seek to entangle novel quantum systems. Many inequalities exist whose violation provides evidence against the proposition of local-realism (under reasonable assumptions) and thus shows quantum behavior.\\

\noindent These inequalities require either dichotomic (that is, two-outcome) measurements on correlations (coincidence counts, photonically), or the use of local measurements (singles, photonically) \cite{bell,chsh,FC}. Examples of dichotomic measurements are spin-up or spin-down, horizontal or vertical polarization, etc. Implementing the two-outcome measurements required by most tests of local-realism can be difficult. The other option is to explicitly include the single count rates, which typically are much higher than coincidences, requiring very high efficiency. For example, in the seminal paper by Aspect \cite{aspect1} only polarizers are used \--- not polarizing beam splitters \--- and in fact it is not until later \cite{aspect2} that true dichotomic measurements are used.\\

\noindent An exception to these conditions is the Wigner Inequality (WI), which needs only single-outcome coincidences \cite{wig}. The price for this flexibility is that the original derivation of this inequality assumes perfect anti-correlations.\\

\noindent Stated most concisely, in its simplest form as derived in 1969, the Wigner inequality gives a restriction on the statistical measurement outcomes of potential hidden variable models (HVM) of perfectly anti-correlated, two-particle, spin-1/2 systems \--- given three potential measurement settings. Specifically, it states that the probabilities of getting coincident detections (either spin-up, spin-up; or spin-down, spin-down) must obey the relation $P_{13}\leq P_{12}+P_{23}$ where the subscripts indicate one of the three settings (typically projection angles) on either the first or second particle (indicated by first or second position in the subscript). This inequality can be violated quantum mechanically with a singlet state by $3/4\leq1/4+1/4$. These basic statements will be derived carefully and expanded upon in what follows.\\

\noindent Modified versions of the WI have been found which do not require perfect anti-correlations, however these again require dichotomic measurements \cite{w2,w3}. In this paper we derive an extended version of the Wigner inequality which does not require dichotomic measurements, use of the singles, or perfect anti-correlation. This allows violation of local realism in a much broader class of experimental systems than was possible before. Examples of such systems are long-(or lossy)-baseline quantum protocols, momentum entanglement in Bose-Einstein condensates with micro-channel plate detectors, the high orbital angular momentum entanglement setup we consider in this paper, or any other such system where singles are higher than coincidences and/or two-outcome measurements are difficult to implement.\\

\noindent In the following section an extended version of the original Wigner inequality is derived and we carefully outline the necessary assumptions, including how we maintain the single-outcome-only measurements despite the reality of imperfect anti-correlations. In the third section we apply the newly derived inequality to a recent experiment \cite{us} utilizing very high orbital angular momentum \--- a situation in which other tests of local realism may not be applied (due to the nature of the detection scheme). In the final section we summarize the main points of the paper.

\section{Derivation of the General Inequality}

\noindent Following Wigner \cite{wig}, assume that there exists a deterministic description of the world which relies on a hidden variable \--- which completely defines the results of all possible measurements, but which is not in principle known or knowable. When a measurement is performed, the statistical distribution of the hidden variable reproduces the probabilities for the potential outcomes. We need not be concerned about the actual statistical distribution of the variable, but only define domains of it which correspond to obtaining a given result when a particular measurement is performed. Each of these domains will be represented by a symbol which, herein, will be called a ``Wigner-symbol".\\

\noindent Take the case of a two particle system, where some degree of anti-correlation is present, and assume that these anti-correlations can be explained via classical statistics. Note that throughout this paper, when we say `(anti-)correlation' we mean `(anti-)correlated detection event' \--- or `giving rise to an (anti-)correlated detection event'. Now limit the measurements settings to three possibilities for each particle $\phi_{1}$, $\phi_{2}$, and $\phi_{3}$. We limit the measurement outcomes to two possibilities: a detection, or no detection. So, we can define the Wigner-symbols (representing domains of the space of the statistical distribution of the hidden variable) in the form $(R_{1a},R_{2a},R_{3a};R_{1b},R_{2b},R_{3b})$, where the $R$'s can take the values of ``$+$" or ``$-$", indicating a detection, or no detection, respectively at the setting, and particle ($a$ or $b$), indicated by the subscripts. If we had the case of perfect anti-correlation we could write only symbols which exhibited this quality, like $(++-,--+)$. This symbol would mean, for example, if a $\phi_{1}$ measurement were performed on particle $a$ and a $\phi_{2}$ measurement were performed on particle $b$ the results would be a detection and no detection, respectively. However, we must allow for the possibility of symbols which do not represent perfect anti-correlations, as particle pairs which are not perfectly anti-correlated can and do occur. This generality is an extension to the original inequality as derived by Wigner.\\

\noindent The Wigner-symbols also have the direct interpretation as the probabilities that the hidden variables take a value lying in the domain specified by that symbol. That is, the size of the domain corresponding to the indicated measurement outcomes is the probability of such an outcome occurring. We have made no assumptions about the statistical distribution of the hidden variables. Now, in order to assume that there exists a realistic description of the world, the probabilities for some chosen outcomes of the described setup can be written as

\begin{eqnarray}
P_{13}&=&(++-,--+)+(+--,-++)+\mathcal{S}_{13}, \label{E1} \\
P_{12}&=&(+-+,-+-)+(+--,-++)+\mathcal{S}_{12}, \label{2}\\
P_{23}&=&(-+-,+-+)+(++-,--+)+\mathcal{S}_{23}. \label{3}
\end{eqnarray}

\noindent Where the $P_{ij}$'s are the probabilities of getting a coincident detection when the detectors are set to the measurement positions indicated by the subscripts. For example, $P_{13}$ is the probability that we detect particles $a$ and $b$ if the measuring devices are in settings $\phi_{1}$ and $\phi_{3}$ respectively. The symbols $\mathcal{S}_{ij}$ represent a collection of those Wigner-symbols which do not exhibit the perfect anti-correlation, for example $(+++,--+)$. There are 14 such Wigner-symbols included in each $\mathcal{S}_{ij}$. Putting Eqs.(\ref{2}) and (\ref{3}) in a slightly different form

\begin{eqnarray}
P_{12}-(+-+,-+-)-\mathcal{S}_{12}&=&(+--,-++), \\
P_{23}-(-+-,+-+)-\mathcal{S}_{23}&=&(++-,--+).
\end{eqnarray}

\noindent And then using this, we write

\begin{eqnarray}
P_{12}-\mathcal{S}_{12}'&\geq&P_{12}-(+-+,-+-)-\mathcal{S}_{12},\nonumber\\
&\geq&(+--,-++),\label{in1}\\
P_{23}-\mathcal{S}_{23}'&\geq&P_{23}-(-+-,+-+)-\mathcal{S}_{23},\nonumber\\
&\geq&(++-,--+).\label{in2}
\end{eqnarray}

\noindent Where $\mathcal{S}_{12}'$ ($\mathcal{S}_{23}'$) consists of only those Wigner-symbols representing imperfect anti-correlations for which the same symbol exists in both $\mathcal{S}_{12}$ ($\mathcal{S}_{23}$) and $\mathcal{S}_{13}$. Direct substitution of Eq.(\ref{in1}) and Eq.(\ref{in2}) into Eq.(\ref{E1}) yields $P_{13}-\mathcal{S}\leq P_{12}+P_{23}\label{probpre}$. Where $\mathcal{S}$ collects all those Wigner-symbols which exhibit imperfect anti-correlations which can not be canceled out. If we were to take $\mathcal{S}$ to zero we would recover the original result by Wigner. Now,

\begin{eqnarray}
\mathcal{S}&=&\mathcal{S}_{13}-\mathcal{S}_{12}'-\mathcal{S}_{23}'\nonumber\\
&=&(+--,--+)+(+-+,--+)\nonumber \\
& &+(+--,+-+)+(+-+,+-+).\label{E}
\end{eqnarray}

\noindent The first line explicitly includes all the symbols lacking from Wigner's original derivation. There are 28 terms in total, of which 24 cancel out. This cancellation is possible because there are no additional ``labels'' on the symbols due to outcomes on the other side \--- that is, due to locality. In the second line we write the result. The second and third terms can be bounded from above by observing the coincident intensity at a measurement setting where there is a minimum \--- at $\phi_{a3}$ and $\phi_{b3}$ for the second term, and $\phi_{a1}$ and $\phi_{b1}$ for the third. Either of these measurements will also set an upper-bound on the fourth term. More problematic is the first term, if dichotomic measurements were available an upper-bound could be placed on this term directly, by making measurements of ``no-click" outcomes.  However, in their absence, this term will be estimated via an ``extended fairness assumption'' (EFA).\\

\noindent The assumption of fairness of measurement results relates to a loophole in Bell and CHSH-type inequalities, known formally as ``fair sampling''. This loophole can be closed by increasing the detector efficiency to a suitable degree \cite{mar,other}. Our assumption of extended fairness differs. It is however philosophically reminiscent, in the sense that it is in some way an assumption that the hidden variable model is not ``conspiring''.\\

\noindent Precisely, our assumption is that: An imperfect correlation is not more likely to occur when detectors are set to any particular measurement position. That is, a Wigner symbol (representing a counterfactual pattern of potential detection events) which would cause some of those detection events to not be anti-correlated, given a particular measurement choice $\phi_{x}$, is not more likely than one that would cause a non-anti-correlated event given a different measurement $\phi_{y}$ (where $x$ and $y$ can be positions $1$, $2$, or $3$). We define an imperfect correlation as a deviation from perfectly anti-correlated Wigner symbols by one of two classes of processes. The first class is composed of events that would cause there to be no detection, when otherwise there would be one, for example the symbol representing the system goes from $(+--,-++)$ to $(+--,-+-)$, i.e. in the final position of the symbol the $+$ is replaced with a $-$. The second class is composed of events that could cause there to be a detection when otherwise there would be none, as an example we take the reverse of the previous example and the symbol representing the system goes from $(+--,-++)$ to $(+--,+++)$, i.e. in the fourth position of the symbol the $-$ is replaced with a $+$.  We will call the first class `loss' and the second class `gain'. \\

\noindent We would argue that our assumption can be seen as connected to fair sampling \--- as fair sampling assumes that the HVM allows the detection device to fairly sample the ensemble \--- whereas extended fairness assumes that the detection device, and whatever object allows projections in the Hilbert space (e.g. being a polarizer) are both acting fairly and not influencing or influenced by the HVM. However, where the line is drawn between ``detector'' and ``system'' is somewhat arbitrary.\\

\noindent Some reasonable questions that could be asked about this are ``How justified is this assumption?'' and ``Does it restrict the source in any way?''. For a detailed discussion please see the first argument in the appendix (the second argument is best read after the derivation of the inequality). It indicates the answers to these questions are ``well justified'' and ``very little''.\\

\noindent Despite those arguments our statement of extended fairness still exists at the level of an assumption that one could choose to either accept, or not accept, for philosophical reasons. Thus the class of local-realistic theories disproved by violation of the inequality we here derive is restricted to a smaller subset than some others.\\

\noindent It is important to point out that our assumption is different than the ``no-enhancement assumption" of Clauser and Horne \cite{CH}. The no-enhancement assumption states that the addition of a polarizer can not enhance the probability of a detection. Our assumption allows for the polarizer (or other device) to enhance, degrade, or have no effect on counts. Furthermore a Bell-type inequality derived using no-enhancement requires correlation measurements with the polarizers (or other devices) removed, which can create problems in implementation (physical difficulty, and skewed count rates).\\

\noindent Now we move to the application of the assumption in our case. We take the first term of Eq.({\ref{E}}) and determine which symbols, representing perfect anti-correlations, could give rise to the $\mathcal{S}$ term of interest under some physical process (dark count, scattering, etc.). Both $(+--,-++)$ and $(++-,--+)$ could give rise to $(+--,--+)$ after a single physical event (after the loss deviation, as defined previously). Mathematically 

\begin{eqnarray}
(+--,-++)\rightarrow(+--,--+),\\
(++-,--+)\rightarrow(+--,--+).
\end{eqnarray}

\noindent This demonstrates how symbol $(+--,--+)$ is one ``step'' away from one of two perfectly anti-correlated symbols. Now, for each of these two symbols representing a perfect anti-correlation there are six symbols which represent a single event occurring to that symbol. That is, 

\begin{eqnarray}
& &(+--,-++)\nonumber\\
& & \quad \rightarrow\left\{(---,-++), (++-,-++),\right. \nonumber \\
& &\quad \quad \, \, \, \, \, (+-+,-++), (+--,+++), \nonumber \\
& &\quad \quad \, \, \, \, \, \left.(+--,--+), (+--,-+-)\right\}.\label{subss}\\
& &(++-,--+)\nonumber\\
& & \quad \rightarrow\left\{(-+-,--+), (+--,--+),\right. \nonumber \\
& &\quad \quad \, \, \, \, \, (+++,--+), (++-,+-+), \nonumber \\
& &\quad \quad \, \, \, \, \, \left.(++-,-++), (++-,---)\right\}.\label{subs}
\end{eqnarray}

\noindent These collections of symbols contain both those generated by gains and losses. There are three possibilities: 1.) Loss and gain are about equally likely. 2.) Gain is more likely than loss. 3.) Loss is more likely than gain. In the first case all the symbols in Eqs.(\ref{subss},\ref{subs}) are equal and we apply the assumption by saying that since the deviation from a perfect anti-correlation is not more likely to take place at any particular measurement position, then all the symbols on the right hand side of Eqs.(\ref{subss},\ref{subs}) are equally likely and we therefore take $(+--,--+)$ and replace it with either $(+--,+++)$ or $(++-,+-+)$. In the second case such a substitution is actually pessimistic as $(+--,--+)<(+--,+++)$ or $(++-,+-+)$. In the third case we can use $(+--,--+)=(+--,-+-)$ or $(-+-,--+)$, as a loss is just as likely at position 6 (or 1 for the second set of symbols) as it is at position 5 (or 2); again, by application of the extended fairness assumption. In this last case either symbol cancels with one of the (previously unused) symbols in $\mathcal{S}_{12}$ or $\mathcal{S}_{23}$ from Eq.(\ref{2}). That is, for example, the symbol $(+--,-+-)$ contributes to $P_{12}$ but appears nowhere else. Since it is to the right-hand side of the $\leq$ and negative it can be ignored, unless we ``need'' it to cancel the aforementioned substitution. Likewise for the other potential symbol. Thus, case one is the most pessimistic in the sense of giving the HVM the most power. So we shall proceed assuming it is true. \\

\noindent Note that we consider symbols that are only ``one step'' away from perfect anti-correlation when applying the EFA since after more steps the number of symbols which can be produced increases dramatically as the number of steps increases (for three steps there are $6^{3}=216$ different possibilities, many more than the actual number of possible symbols). Thus, it is very unlikely that any anti-correlation will be witnessed by the experimenter at all in such a situation. Furthermore it is simply not necessary to examine \emph{every possible} equivalence between symbols, merely one is sufficient to allow bounding.\\    

\noindent Via the above arguments we now take $(+--,--+)$ and take replace it with either $(+--,+++)$ or $(++-,+-+)$. Similarly we can replace $(+-+,--+)$ with $(+--,++-)$ (this second replacement is not strictly necessary, though it simplifies our derivation without loss of generality); Eq.(\ref{E}) thus becomes either

\begin{flalign}
\mathcal{S}&=(+--,+++)+(+--,++-)&&\nonumber \\
& \quad +(+--,+-+)+(+-+,+-+)\leq P_{11}.
\end{flalign}

\noindent Or,

\begin{flalign}
\mathcal{S}&=(++-,+-+)+(+--,++-)&&\nonumber \\
& \quad +(+--,+-+)+(+-+,+-+)\leq P_{11}.
\end{flalign}

\noindent Where we have once again employed a straightforward relationship between the Wigner-symbols and the probabilities of seeing correlations at specific measurement positions. In both cases the result is the same. Therefore, the inequality now becomes

\begin{eqnarray}
P_{13}-P_{11}\leq P_{12}+P_{23}.\label{prob}
\end{eqnarray}

\noindent Note now that the inequality is written totally in terms of measurable quantities, requiring only single outcome measurements on a bipartite system which displays some degree of anti-correlation. We can re-write it in terms of intensities as

\begin{eqnarray}
I_{13}-I_{\mathrm{min}}\leq I_{12}+I_{23}.\label{final}
\end{eqnarray}

\noindent Where the $I$'s indicate intensities. This constitutes the main theoretical result of this paper. Violation of this inequality negates the possibility of a local-realistic description of the world under the outlined assumptions.\\

\noindent Now, if we are to compare this with a system described by quantum physics, for example, an entangled pair of spin-1/2 particles, described by the pure singlet state, and taking the measurements to be positive projections onto a chosen direction given by angle $\theta$, we have $P_{xy}(\theta_{x},\theta_{y})=\sin^{2}(\theta_{x}-\theta_{y})$.\\

\noindent If we make the choices $\theta_{1}=0^{\circ}$, $\theta_{2}=30^{\circ}$, and $\theta_{3}=60^{\circ}$, then we find: $3/4\leq 1/4+1/4$. So clearly, quantum mechanics contradicts the local-realistic description of the world assumed in the formulation of the Wigner inequality. The chosen angles offer the maximal violation. For this example using pure, maximally entangled singlet states we had $P_{11}=0$. Our inequality allows generalizations to when this is not the case. The theoretical and experimental characterization of one such case will be the subject of the next section.\\

\noindent Before doing so we will revisit the extended fairness assumption. Despite the first argument given in the appendix, the extended fairness does involve a degree of confidence that the experimental setup is performing in a way which is consistent with the assumption. Arguing solely from the classical point-of-view such confidence could be arrived at by observation of the rotational symmetry of the measurement apparatus. For example, for the case of a detector apparatus consisting of a polarizer (or a ``slit wheel'', a device which we will discuss later) and avalanche photo-diode, the assumption could be said to be well-formed if, as the assembly of devices is rotated through its full range of detector positions no (or little) variation is observed in the singles count rate. For details see the second argument in the appendix.\\

\noindent We can qualitatively summarize the assumption by rephrasing it as: The extended fairness assumption is that in a system displaying a large degree of anti-correlation, deviations away from perfect anti-correlation are homogenous with respect to local choices of measurement positions. This is especially likely to be the case when experimenters observe flatness in the local count rates with respect to measurement choice. \\

\noindent It is also worth briefly noting that our inequality is not vulnerable to the possibility of false violation arising from improperly normalizing a two-outcome measurement which exists in a larger dimensional Hilbert space (see for example Ref.\cite{false}). This due to the fact that in our case all unobserved outcomes are course-grained into ``no coincident detection'' and since the inequality is linear in probability \--- it requires no normalization. In other words there is no danger of assuming that an orthogonal event occurred when in fact there simply was no event at all since we do not make a distinction between these two possibilities.\\

\section{Violation with Very High Orbital Angular Momentum}

\noindent A recently reported experiment on entangling states of very high orbital angular momenta of light \cite{us} presents a difficult case if one wants to violate local realism using the generated entanglement.\\

\noindent For paraxial beams the total angular momentum separates out into well-defined spin \cite{poy} and orbital \cite{A} components. Laguerre-Gauss beams naturally carry integer amounts of orbital angular momentum (OAM) in units of $\hbar$, as the quantum number $l$.  One promising avenue of research is using the OAM degree of freedom as an infinite alphabet for quantum informatics tasks, and quantum communication \cite{QKD,QKD2,wien}. An important element of many such protocols is entanglement of beams with different OAM.\\

\noindent In our experiment, the objective is to generate entanglement of the target form $|l\rangle_{a}|-l\rangle_{b}+|-l\rangle_{a}|l\rangle_{b}$. This is done by initially creating entanglement with a Sagnac source and transferring the polarization entanglement to OAM with an interferometric scheme. For more details see Ref.\cite{us}. The subscripts $a$ and $b$ label which of two spatial modes the photon occupies, and $\pm l$ is the OAM quantum number of the photon.\\

\begin{centering}
\begin{figure}[h]
\includegraphics[scale=0.6]{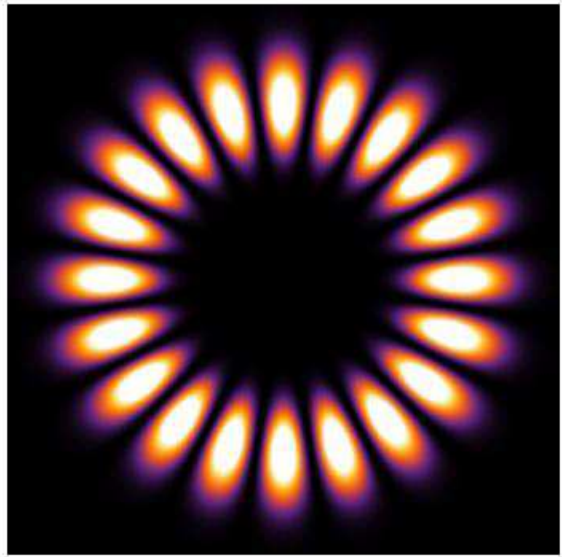}
\caption{The calculated transverse intensity pattern of a supperposition of two LG modes of OAM quantum numbers $10$ and $-10$ (and radial number 0), showing the $20$ azimuthal intensity fringes. These fringes exist in coincidence for entangled states. \label{L}}
\end{figure}
\end{centering}

\noindent This state is maximally entangled in the OAM basis which spans a two-dimensional subspace of the full infinite-dimensional OAM Hilbert space. Another notable feature of this state is that its coincident-intensity, between the two beams, as a function of angular position, displays an interference pattern with $2l$ fringes. One such pattern is shown in Fig.{\ref{L}}.\\

\begin{centering}
\begin{figure}[h]
\includegraphics[scale=8]{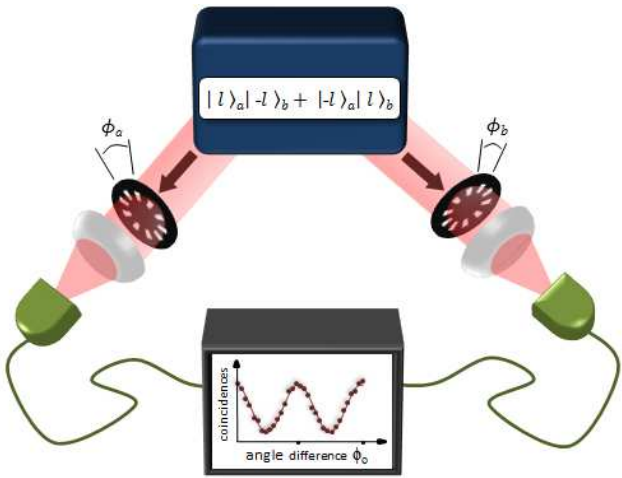}
\caption{The setup produces two beams of entangled photons, which are then incident on slit wheels rotated to the given angles. Behind the slit wheels are bucket detectors. Fringes as a function of slit-wheel-angle difference can be seen in the coincidence counts. \label{1}}
\end{figure}
\end{centering}

\noindent Specifically, the coincident intensity, calculated quantum mechanically, is given by $\cos^{2}(l(\phi_{a}-\phi_{b}))$, where $\phi_{a}$ and $\phi_{b}$ are the angular positions. Given this information, a way of determining whether or not a source is producing the target state presents itself rather intuitively. That is, placing a ``slit wheel'' (a black absorptive disc with rectangular holes cut into it radially) with $2l$ transparent slits in each beam, and then rotating one or the other of them so that the interference pattern can be read out by bucket detectors placed behind them as a function of slit-wheel position. If the pattern matches the expected fringes, then it can be concluded that the target state is being produced. This is illustrated in Fig.{\ref{1}}.\\

\noindent This procedure though simple, intuitive, and inexpensive, is difficult to turn into a test of the classical world-view. Dichotomic measurements are not available because the slit wheel does not project in a two-dimensional space. This means inequalities of the type employed by Aspect in his third experiment \cite{aspect2}, as well as any other dichotomic scheme, may not be employed. Furthermore, due to the very lossy nature of each slit wheel, single counts are always much greater than coincidences. Thus, any inequality which employs both coincidences and singles (such as in Ref.\cite{CH}) will be impossible to violate. However, these are not insurmountable problems for the inequality derived in the previous section, as we do not require dichotomic measurements and the problems caused by loss are less dramatic.\\

\noindent It is worthwhile to mention in passing a recent paper that might be helpful in this situation; based on CSHS-type inequalities for continuous periodic variables: Ref.\cite{con}.\\

\noindent Before proceeding, we note that some of the noise present in the ``compensation term'' ($I_{\mathrm{min}}$), will be the result of the object performing the projection in the Hilbert space of the state being measured. For the case of the slit wheel a significant reduction of visibility is due to the fact that the slits are of finite width. However this effect can be predicted precisely with a theoretical model and then subtracted off from the results.\\

\noindent First, we write the Laguerre-Gauss beam in terms of angular position states

\begin{eqnarray}
|l\rangle=\frac{1}{\sqrt{2\pi}}\int^{2\pi}_{0}d\phi e^{il\phi}|\phi\rangle.
\end{eqnarray}

\noindent The phase factor comes from the transverse phase profile of the LG modes. Now we can also write the projection operators representing the slit-wheels and bucket detectors

\begin{eqnarray}
\hat{O_{a}}&=&\sum_{n=0}^{2l-1}\int^{\frac{\pi}{l}(n+W_{s})}_\frac{\pi n}{l}d\phi_{a}'|\phi_{a}'\rangle\langle\phi_{a}'|.\\
\hat{O_{b}}&=&\sum_{n=0}^{2l-1}\int^{\frac{\pi}{l}(n+W_{s})+\phi_{o}}_{\frac{\pi n}{l}+\phi_{o}}d\phi_{b}'\left|\phi_{b}'-\frac{\pi}{2l}\right\rangle\left\langle\phi_{b}'-\frac{\pi}{2l}\right|.\nonumber
\end{eqnarray}

\noindent Where $W_{s}$ is the slit width as a fraction of the distance to the next slit, and $a$ and $b$ label which beam the wheel is placed in. The shift in the projector states by $\pi/2l$ indicates that the data (or wheel) should be shifted one half-cycle. This relabeling causes the events to be anti-correlated as opposed to correlated. The factor $\phi_{o}$ represents the relative difference in angle between the slit wheels. We can then write the expectation value of both slit wheel operations, $\langle\hat{O}_{a}\hat{O}_{b}\rangle$, as

\begin{eqnarray}
P(\phi_{a},\phi_{b})=W_{s}^{2}-\frac{\cos(2l\phi_{o})}{\pi^{2}}\sin^{2}(\pi W_{s}).\label{prob}
\end{eqnarray}

\noindent We can now make the replacement in Eq.(\ref{final})

\begin{eqnarray}
I_{\mathrm{min}}\rightarrow I_{\mathrm{max}}\left(\frac{I_{\mathrm{min}}}{I_{\mathrm{max}}}-\frac{P_{\mathrm{min}}}{P_{\mathrm{max}}}\right).
\end{eqnarray}

\noindent Where the $I$'s are the experimentally measured values and the $P$'s are the predicted probabilities. This compensates for the finite slit width by subtracting off the noise we have theoretically characterized as not originating from inaccuracies in the state.\\

\noindent For the case of the state $|100\rangle_{a}|-100\rangle_{b}+|-100\rangle_{a}|100\rangle_{b}$ we choose the measurement angles $\phi_{1}=0^{\circ}$, $\phi_{2}=0.3^{\circ}$, and $\phi_{3}=0.6^{\circ}$ and find $I_{13}=5654\pm 75$, $I_{\mathrm{min}}=991\pm 31$, $I_{\mathrm{max}}=7845\pm 89$, $I_{12}=2202\pm 47$, and $I_{23}=2456\pm 50$ in terms of total counts, assuming possionian statistics. The slit width used is $W_{s}=0.149$, yielding $P_{\mathrm{min}}=0.002$, and $P_{\mathrm{max}}=0.043$ This results in a violation of Eq.(\ref{final}) by $368\pm 135\leq 0$, a violation with a confidence of a little over two sigma. This experiment represents the first violation of local realism with such high quanta.\\

\noindent It is useful to make a few short comments regarding these results. In other tests of local realism where two-outcome measurements are not assumed (e.g. being Ref.\cite{CH}) it is necessary to subtract singles and coincidences. The singles count rates for this experiment are on the order of $10^7$ whereas the coincidences are of the order of $10^4$. Thus violation would never be possible using other inequalities. The principle aim of this experiment is two-fold: Firstly, to violate local realism in a very novel setup, and thus prove the existence of entanglement via a different method than in the original paper. Secondly, to show the robustness of our derived inequality in very difficult conditions; showing that the inequality has utility in situations where similar difficulties exist that is: coincidences much lower than singles, and difficultly (or impossibility) of performing two-outcome measurements. Much better experiments exist whose express purpose is to disprove local-realism as a valid description of the world and close as many loopholes simultaneously as possible. However this is not the aim of this paper.\\

\noindent It is important to remark that a high degree of flatness is observed in the singles count rate for both detector stations. As we argued in the text and appendix this is good qualitative evidence that the assumption of extended fairness is valid in this particular experimental setup.\\  

\noindent A fair question would be ``Why is the `compensation' due to finite slit width applied only to the term $I_{\mathrm{min}}$ and not to the other three terms?'' In response we would point out that as the difference angle between the two slit wheels approaches it's intensity maximum ($\phi_{o}=\pi/4l$) the effect from the finite width of the slit wheels contributes less noise to the correlations. To see this take Eq.(\ref{prob}): The second term is the one that couples slit width to angular position. It is smaller nearer a maximum. Thus if we were to apply similar corrections to the other terms the effect would be significantly smaller. Furthermore, since the effect would be smallest for the $C_{13}$ term, applying the correction to all terms would only help us achieve violation more easily (though, only slightly more easily). Thus, not doing the correction on the other terms is the more conservative approach.

\section{Conclusions and Outlook}

\noindent We have derived a new version of the Wigner Inequality which can be applied to any experimental setup where some degree of anti-correlation is present between two particles. Violation of this inequality excludes the possibility of a local-realistic description of the experiment, under a set of assumptions which we have outlined. Unlike Bell and CHSH-type inequalities, the derived inequality does not require two-outcome measurements or use of singles \--- allowing its application to systems to which other tests of local realism can not be successfully applied. We experimentally violated the inequality in a novel system: an experiment where very high orbital angular momenta are entangled. Due to the particulars of the detection scheme, other local-realistic inequalities could not be violated. We believe that the derived inequality to be of high utility, not merely to the specific system we apply it to, but also to a wide range of experiments where applying other inequalities may be difficult, if not impossible.

\section{Acknowledgements}

\noindent The authors would like to thank Anton Zeilinger for initiating and motivating this work, and for suggesting that the Wigner inequality had potential in real experiments. We would also like to thank J\'{a}nos A. Bergou, Johannes Kofler, Mario Krenn, and Marcus Huber for useful discussions. This work was supported by the ERC Advanced Grant QIT4QAD, and the Austrian Science Fund FWF within the SFB F40 (FoQuS) and W1210-2 (CoQuS). S.R. is supported by an EU Marie-Curie Fellowship (PIOF-GA-2012-329851).

\section*{Appendix: The Extended Fairness Assumption}

\noindent One way we can attempt to convince the skeptical reader that our assumption is well justified is by arguing how unlikely it is for the assumption to be invalid in an experimental setup which displays \emph{flat singles}. Below we present two arguments to this effect, from two different perspectives. \\

\subsection*{First Argument} 

\noindent Since the inequality we derived is linear, due to convexity, the maximum violation is for a pure state \cite{marky}. For entangled states (those which can violate the inequality) which are pure, or close to pure, the local measurements (i.e. the singles) show very little to no variation as detector angles are continuously varied (if efficiency is independent of detector orientation). This can be understood from the fact that the partial trace over states which are highly entangled results in local states which are highly mixed. In the case that the \emph{local} states are highly mixed prior to interaction with the detection apparatus, extended fairness becomes analogous to the, more traditional, fair sampling assumption. To continue the qualitative argument, the extended fairness assumption can be seen as most likely valid when flatness is observed in the singles. There are of course states whose statistics are not flat, locally, but these are those which will be least likely to result in violation, as we explain above.\\

\subsection*{Second Argument}

\noindent To continue, to see why flat singles are strong evidence for the accuracy of our assumption from the classical perspective, consider the case where such an observation of flat singles is made and yet the hidden variable model \emph{can} disregard the assumption: An implication of the assumption (in fact the one that is used in the derivation) is that $(+--,--+)\approx (+--,+++)$ \--- interpreting these symbols as being probabilities with numerical values (as stated in the main text this is the most pessimistic possibility in terms of giving the HVM the most power, so it is what we assume). Consider a HVM which can ignore this restriction and cause $(+--,--+)$ to be larger, thus violating the inequality while maintaining classical realism (In the extreme case our inequality can be trivially violated by a HVM which produces \emph{only} systems defined by this symbol and no others). A HVM with the best chance of ``tricking'' an experimenter would start with an evenly weighted statistical mixture of perfectly anti-correlated classical states (i.e. described by the symbols which appear in the original derivation by Wigner; the even mixture resulting in flat singles), and adding enough $(+--,--+)$ to violate the inequality. But now Alice and Bob would see their singles spike at settings one and three respectively. To compensate for this, while still having a violation, the HVM could now add more pairs described by symbols such as $(--+,+--)$ and $(-+-,-+-)$. But now the HVM is attempting a ``balancing act''. Furthermore such a malicious HVM would have to produce many symbols which are not close to being anti-correlated, a situation which would be obvious to an experimenter if he or she makes measurements with the settings the same for Alice and Bob's stations. For certain, a conscious conspiracy could engineer such a situation (or use a similar, though equally difficult strategy) and keep it hidden from the experimenter (as it is likewise for inequalities using the more familiar fair-sampling assumption), but we argue it's reasonable to assume that it is very unlikely to happen by chance.\\

\noindent We will now attempt to quantify the phrase ``very unlikely'' with some toy models of experiments describable with HVMs. First take a physical system described only by the perfectly anti-correlated Wigner symbols ($8$ in total). Now assume that in this model that symbols can only be either ``on'' or ``off''. If a symbol is off it has probability zero. If it is on it has a probability equal to every other symbol which is on. Now we require that the HVM produce measurement statistics such that the singles count rates are flat. By inspection there $25$ such combinations of on/off for the perfect symbols. There are $2^{8}=256$ total combinations of on/off when no requirement is made of the singles. Thus, if we are are to ask the question ``What are the chances of a randomly chosen perfectly classically anti-correlated system displaying flat singles?'' The answer is ``Roughly, $1$ in $10$''.\\

\noindent Extension of this logic to \emph{imperfectly} anti-correlated HVMs which obey the EFA is straightforward. Consider symbols that are ``one step'' away from being perfectly anti-correlated. There are $24$ such symbols. If we take the case of the EFA being true then certain groups of these symbols must be all on, or all off. The groups are such that the same proportion of combinations of these new symbols display flat singles.  Put another way, if a combination of perfectly anti-correlated symbols displays singles with some degree of flatness then the imperfectly anti-correlated symbols that ``devolve'' from these combinations must have the same degree of flatness since each position in the symbol is fairly switched from $+(-)$ to $-(+)$. Thus, as with the perfectly anti-correlated symbols, the chances of a random non-perfectly anti-correlated HVM obeying EFA having flat singles is roughly $1$ in $10$.\\

\noindent Now we turn our attention to HVMs which have \emph{no restriction} on which symbols may be jointly off or on. The number of free, unique combinations of the $24$ symbols which lead to flat singles may be found via combinatorical methods and is $2^{12}-13=4083$. However the \emph{total} number of possible symbols is $2^{24}$. So the probability of a randomly chosen HVM \--- without the EFA \--- arriving at flat singles is roughly $1$ in $4100$.\\

\noindent Now if an experimenter observes flat singles in his or her experiment and asks the question ``What are the chances this is due to a HVM operating consistently with EFA; compared to a model which has no such restriction?'' The answer would be  ``Roughly $400$ times more likely.'' This toy model rests on some simplifications \--- perfect flatness, symbols being only ``off'' or ``on'', and the idea that HVMs be chosen randomly from equivalent levels of anti-correlation. Thus, these models should not be used to make concrete statements but serve as \--- we conjecture \--- strong qualitative evidence that in systems where a degree of anti-correlation and flat singles are observed the extended fairness assumption is a safe one.

\end{document}